# Consumer lying in online reviews: recent evidence

**Shawn Berry, DBA[1]***

**May 20, 2024**

[1]William Howard Taft University, Lakewood CO, USA

*Correspondence: shawn.berry.8826@taftu.edu

**Abstract**  The persistence of lying by some consumers in their online posts of experiences with businesses is problematic, and taints the global pool of information that is used for decision making by people that assume they are true accounts of experiences. This study is based on data from my dissertation about fake online Google reviews of restaurants (Berry, 2024), and leverages an instrument that quantifies the trust of people. The findings are based on a sample of n=351, and provide a general proxy for lying in online reviews, and sketch out the characteristics of a typical person that has the propensity to be untruthful. A predictive model of posting untrue online reviews is constructed. The findings have wider implications for the study and monitoring of deceptive behavior, including the propagation of misinformation, and a means of quantifying the potential for antisocial behavior as measured by the trust of people instrument in Berry (2024). Directions for future research and limitations are also discussed.

## 1. Introduction

Several scholars have looked at the phenomenon of lying by consumers in online reviews (Drouin, Wehle & Hernandez (2016), Ott, Cardie & Hancock (2012), Román, Riquelme, & Iacobucci (2019), Petrescu et al. (2022)). The behavior of lying has several implications for consumers and the businesses that they are lying about with respect to the quality of a shopping experience or satisfaction with a product or service before or after a transaction. Untruthful reviews are problematic for businesses because they deliberately misrepresent the account of the experience to the extent that it creates an unwanted bias in the decision making process for other consumers.

Studies by several scholars into the extent that untruths are used in online reviews suggests the percentage of reviews that are not true varies from 16% and 33% (Luca and Zervas, 2016; Munzel, 2016; Salehi-Estafani and Ozturk, 2018; Schuckert, Liu and Lau, 2016; Wu et al., 2020). With respect to the frequency of lying as a behavior among the general population in the United States, Serota, Levine and Boster (2010) estimated "that on average Americans tell one to two lies per day" (p.8), and note that "sixty percent of subjects report telling no lies at all, and almost half of all lies are told by only 5% of subjects; thus, prevalence varies widely and most reported lies are told by a few prolific liars" (p.1). Verigin et al. (2019) also found "participants indicated telling a mean of 1.61 lies during the last 24 hours" (p.4). These observations imply that while the contingent of liars that post untrue reviews online are relatively few, the estimates of the prevalence of the behavior are nontrivial and cause for concern as to the potential to deceive and misinform consumers.

Although my dissertation did not explore the reasons why respondents chose to post untrue online reviews, Kapoor et al. (2021) suggests that lying by consumers can be characterized as exaggeration of an experience (whether good or bad) and is perhaps the product of an underlying personality defect in a person such as narcissism or psychopathy, for example, that creates a deviation away from good moral principles. While this is not a singular reason for a person to lie, the observation by Kapoor et al. (2021) implies that lying behavior is perhaps rooted in a possible tendency for people to exploit a situation or other people if they generally lack fundamental moral standards. Another explanation is offered by Levine and Duncan (2022) in their discussion of the effects of deception in the communication of ideas, stating that "deception is seen as the most justified when it is morally motivated and when it involves indirect tactics that are not perceived as particularly dishonest" (p.33). Zimbler and Feldman (2011), in their study of the contextual role of online lying behavior, reduce the problem by stating "that it may be normative to distort reality online" (p.2492).

Empirical studies of dishonest behavior in the field of psychology have attempted to link the propensity to lie with various moderating variables. Gerlach, Teodorescu and Hertwig (2019) observe that "with respect to age effects, some studies have found that younger participants behave more dishonestly than older participants" (p.5), and note that little is understood in the scholarly literature about what underlies this factor. Gerlach, Teodorescu and Hertwig (2019) also observe that the effect of gender on lying behavior is also less understood because "on a broader scale, empirical evidence on gender effects is not clear cut" (p.4) since many studies do not consistently conclude that one gender is always more prone to lying than another. Rather, the authors imply that although some studies find men to take risks and lie more than women, the reasons for lying are largely related to the circumstances and some perceived balance of reward and risk. Although Abeler et al. (2014) suggest that "income could be positively correlated with honesty because of the lower marginal utility of the monetary rewards or negatively correlated because of reverse causality" (p.99), Piff et al. (2012) suggest that "upper-class individuals behave more unethically than lower-class individuals" (p.4086). In their study on cheating behavior, Ariely et al. (2019) found that "higher levels of education reduce the probability of cheating" (p.184), and "that age has a significant impact on cheating" (p.184). Finally, the trust of people is a factor in the extent to which someone is given credibility. Brühlmann et al. (2020) point out the link between how trust is endowed to strangers online when consumers making a decision, observing that "trust is an essential factor in many social interactions involving uncertainty. In the context of online services and websites, the problems of anonymity and lack of control make trust a vital element for successful e-commerce" (p.29).

This paper will examine the correlates of consumer lying behavior with respect to demographic attributes to determine which variables are predictors of the propensity of a consumer to lie in an online review. This insight is valuable because it can also imply the general propensity for people to engage in other morally disengaged behaviors that leverage lying as a means to achieve a goal, such as the propagation of misinformation and being untruthful in transactions or high-stakes negotiations of some kind, for example. Additionally, the trust of people as a variable will be examined as a predictor of lying behavior in online reviews. The study will present a predictive model of lying behavior in online reviews. Finally, the study will conclude with implications for future research.

## 2. Materials and Methods

The data for this paper comes from my dissertation on fake online Google restaurant reviews, wherein an online questionnaire was administered by way of Amazon Mechanical Turk (mTurk) (Berry, 2024). Some of the questions were designed to collect information about beliefs and habits of respondents, including demographic variables. Among the questions posed about online review posting habits, respondents were asked whether they had ever posted an online review that was untrue.

The survey panel size was a convenience sample of n=398 male and female mTurk users from the United States (Berry, 2024). However, the final sample size was reduced to n=351 after 11.8% of the 398 respondents were excluded due to not consenting to the study (3), self-reporting as not residing in the United States (8), attention check question failure (24), and identification of attempts to take the survey more than once (12) (Berry, 2024).

The trust of people construct used in Berry (2024) required respondents to complete an instrument within the questionnaire wherein a set of 13 individuals must be rated according to the extent to which they trust or distrust using a 5-point Likert scale, and quantified by adding up the scores. Each individual in the set represented different people from various walks of life, job roles, and levels of authority, from very personally familiar and close, such as friends and family, for example, to not so personally familiar, such as celebrities and strangers on the street, for example. Using Cronbach's alpha, the reliability of the instrument was 0.85 (Berry, 2024), which is considered to be high (Taber, 2018), and therefore, reliable.

The data was coded for analysis according to the scheme as illustrated in Table 1 below.

**Table** 1

*Variable coding scheme*

| Variable name | Variable description | Variable type | Coding |
|---|---|---|---|
| Age | Age group of respondent, years of age | Categorical | 18-24 = 1<br>25-34 = 2<br>35-44 = 3<br>45-54 = 4<br>55 and over = 5 |
| Gender | Gender of respondent | Categorical | Female = 0<br>Male = 1<br>Non-binary = 2 |
| Income | Annual income level of respondent, USD | Categorical | Less than $30,000 = 1<br>$30,000-$49,999 = 2<br>$50,000-$69,999 = 3<br>$70,000 and over = 4 |
| Education | Education level of respondent | Categorical | Did not finish high school = 0<br>High school graduate = 1<br>Some college = 2<br>Bachelor's degree = 3<br>Master's degree = 4<br>Post-graduate or higher = 5 |
| Region | United States region of | Categorical | Middle Atlantic = 1<br>New England = 2<br>South Atlantic = 3 |
| | residence of respondent | | East South Central = 4<br>West South Central = 5<br>Mountain = 6<br>Pacific = 7 |
| Likert scores | Degrees of agreement or importance of behavioral factors to measure trust of people | Ordinal | Definitely distrust = 1<br>Somewhat distrust = 2<br>Neither trust nor distrust= 3<br>Somewhat trust = 4<br>Definitely trust = 5 |

**Source:** Adapted from Berry (2024), Table 18.

## 3. Results

The sample data was analyzed and modeled. Table 2 illustrates the number of respondents that declared whether or not they ever posted an online review that was untrue (Berry, 2024). These findings suggest that while the majority of respondents claim to be truthful, 16.5% of respondents admit to having posted online reviews that were not true (Berry, 2024).

**Table 2**

*Respondents That Have Ever Posted an Untrue Online Review*

| Have you ever posted an online review that was not true? | *N* | % of total |
|---|---|---|
| No | 293 | 83.5% |
| Yes | 58 | 16.5% |
| Total | 351 | 100.0% |

**Source:** Berry (2024), Table 16.

Table 3 illustrates the number of respondents according to their gender that declared whether or not they ever posted an online review that was untrue. While females represented the majority of the total sample (68.4%), of the 58 total respondents that admitted to posting untrue reviews, 56.9% were female respondents, or 9.4% of the total sample. Among those respondents that did not post untrue reviews, female respondents represented the majority of this set of respondents, 70.6% or about 59% of the total sample. Of all males, 23.1% admit to posting untrue online reviews (25 of 108), and about 14.8% of all females admit to having done so as well.

**Table 3**

*Frequency of respondents that admitted to posting reviews that were untrue according to gender*

| Gender | No | Yes | Grand Total |
|---|---|---|---|
| Female | 207 | 33 | 240 |
| Male | 83 | 25 | 108 |
| Non-binary | 3 | 0 | 3 |
| Grand Total | 293 | 58 | 351 |

**Source:** Data analysis, Berry (2024).

Table 4 below illustrates the number of respondents that declared whether or not they posted an online review that was not true according to their region of residence in the United States. The highest number of respondents that posted truthful reviews was in the South Atlantic region (82 of 293, 23.4% of the total sample). The highest number of respondents that posted untruthful reviews was in the Middle Atlantic region (22 of 58, 6.3 % of the total sample). The majority of truthful respondents were concentrated among the Atlantic regions and South Central regions. The lowest number of respondents that posted untruthful reviews was in New England.

**Table 4**

*Frequency of respondents that admitted to posting reviews that were untrue according to region of United States*

| Region | No | Yes | Grand Total |
|---|---|---|---|
| Middle Atlantic (NY/NJ/PA) | 45 | 22 | 67 |
| New England (CT/ME/MA/NH/RI//VT) | 15 | 1 | 16 |
| South Atlantic (DE/DC/FL/GA/MD/NC/SC/VA/WV) | 82 | 8 | 90 |
| East South Central (AL/KY/MS/TN) | 51 | 6 | 57 |
| West South Central (AR/LA/OK/TX) | 42 | 8 | 50 |
| Mountain (AZ/CO/ID/MT/NV/NM/UT/WY) | 24 | 4 | 28 |
| Pacific (AK/CA/HI/OR/WA) | 34 | 9 | 43 |
| Grand Total | 293 | 58 | 351 |

**Source:** Data analysis, Berry (2024).

Table 5 below illustrates the number of respondents by level of education that declared whether they posted an online review that was not true. The majority of respondents that declared not having posted an untrue online review were those that possessed some college or a bachelor's degree (162 of 293, 46% of the total sample). The majority of respondents that declared to have posted an untrue online review were among those possessing bachelor's and master's degrees (44 of 58, 12.5% of the total sample). Respondents with postgraduate degrees or only a high school education were the least likely to post an untrue online review.

**Table 5**

*Frequency of respondents that admitted to posting reviews that were untrue according to level of education*

| Education level | No | Yes | Grand Total |
|---|---|---|---|
| Did not finish high school | 2 | 0 | 2 |
| High school graduate | 43 | 3 | 46 |
| Some college | 119 | 9 | 128 |
| Bachelor's degree | 92 | 20 | 112 |
| Master's degree | 25 | 24 | 49 |
| Postgraduate or higher | 12 | 2 | 14 |
| Grand Total | 293 | 58 | 351 |

**Source:** Data analysis, Berry (2024).

Table 6 below illustrates the number of respondents by age category that declared whether or not they posted an online review that was not true. Coincidentally, the 25 to 34 year and 35 to 44 year age categories comprised the majority of truthful (209 of 293, 59.5% of total sample) and untruthful (44 of 58, 12.5% of total sample) respondents. Respondents aged 55 and older were the least likely to post an online review that was not true, followed by those aged 18 to 24.

**Table 6**

*Frequency of respondents that admitted to posting reviews that were untrue according to age categories*

| Age category | No | Yes | Grand Total |
|---|---|---|---|
| 18–24 | 22 | 4 | 26 |
| 25–34 | 112 | 27 | 139 |
| 35–44 | 97 | 17 | 114 |
| 45–54 | 51 | 8 | 59 |
| 55 and older | 11 | 2 | 13 |
| Grand Total | 293 | 58 | 351 |

**Source:** Data analysis, Berry (2024).

Table 7 below illustrates the number of respondents by income level that declared whether or not they posted an online review that was not true. The highest number of respondents that posted online reviews that were true earned less than $30,000 per year, and those that posted reviews that were not true earned between $30,000 and $49,999 per year. The majority of respondents (79.3%) that posted untrue reviews earned $49,999 or less. A small number of respondents with an annual income of $70,000 or more admit to posting online reviews that were not true. The lowest number of respondents that posted untrue reviews was in the $50,000 to $69,999 level.

**Table 7**

*Frequency of respondents that admitted to posting reviews that were untrue according to level of income*

| Income level | No | Yes | Grand Total |
|---|---|---|---|
| Less than $30,000 | 104 | 22 | 126 |
| $30,000–$49,999 | 82 | 24 | 106 |
| $50,000–$69,999 | 46 | 3 | 49 |
| $70,000 and over | 61 | 9 | 70 |
| Grand Total | 293 | 58 | 351 |

**Source:** Data analysis, Berry (2024)

The extent to which respondents admitted to posting untrue online reviews was evaluated with respect to the levels of trust or distrust certain kinds of people as collected in the trust of people instrument in Berry (2024). Table 8 summarizes these attitudes, grouped by levels of trust according to each kind of individual. While many of these respondents indicated that they are somewhat and/or definitely trust the different roles or kinds of people instead of definitely distrusting, the findings suggest that large proportions of respondents were indifferent, and therefore, a potential expression of distrust. For example, the number of respondents that were indifferent to trusting or distrusting business owners rivaled the number that had higher levels of trust in business owners. Taken together with the number of respondents that generally distrusted business owners, this finding implies that people do not trust business owners. The greatest number of respondents that had distrust was for celebrities, followed by government officials, and salespeople and immigrants tied as the third most distrusted people. Doctors garnered the least number of respondents expressing distrust of them, followed by family and friends. The greatest number of respondents that expressed indifference was for religious leaders, followed by business owners, new immigrants, and government officials tied with social media influencers. Thus, the ambiguity or indifference as to whether a respondent trusts a certain person or not among this group of respondents that admit to posting untrue online reviews might imply ambivalence, and perhaps be a reliable cue to identifying the potential for lying behavior. In fact, Hartwig and Bond (2011) observed that ambivalence is a strong behavioral cue for deception.

**Table 8**

*Frequency of respondents that admitted to posting reviews that were untrue according to trust levels of people*

| Individual | Definitely and somewhat distrust | Neither trust nor distrust | Definitely and somewhat trust | Total |
|---|---|---|---|---|
| Doctors | 5 | 13 | 40 | 58 |
| Celebrities | 15 | 7 | 36 | 58 |
| Social media influencers | 12 | 15 | 31 | 58 |
| People in commercials | 8 | 12 | 38 | 48 |
| Strangers on the street | 12 | 13 | 33 | 58 |

| | | | |
|---|---|---|---|
| Salespeople | 13 | 9 | 36 | 58 |
| Religious leaders | 11 | 22 | 25 | 58 |
| Teaching professionals | 8 | 12 | 38 | 48 |
| Business owners | 12 | 21 | 25 | 58 |
| Government officials | 14 | 15 | 29 | 58 |
| Friends | 9 | 5 | 44 | 58 |
| Family | 7 | 5 | 46 | 58 |
| New immigrants | 13 | 16 | 29 | 58 |

**Source:** Data analysis, Berry (2024).

The trust of people construct from Berry (2024) was used to evaluate if respondents that posted untrue online reviews had different levels of trust from those respondents that did not. The respondents that declared having not posted an online review that was not true had lower levels of trust of people ($M$=37.2, $SD$ = 6.3) whereas those respondents that declared having posted untrue online reviews had higher levels of trust of people ($M$=46.7, $SD$=11.6), and the difference between the means was not equal to zero, $T(351.44)$= -88.141, $p$<0.001.

### 3.3. Regression analysis

The sample data was analyzed using logistic regression analysis. A model of lying behavior was constructed using the dependent variable *everlied*, which is equal to 1 if a respondent admitted to posting an untrue online review, and 0 if the respondent admits to not having posted an untrue online review. The results are shown in Table 9. The Akaike Information Criterion (AIC) for the model is 236.6. The intercept term, level of education, and trust of people score were each highly statistically significant predictors (p<0.001). Age, gender, and level of income were each statistically significant predictors at the 5% level. However, region was not a statistically significant predictor. Level of education appears to increase the log odds of the behavior the most for a one level increase. As age increases, the log odds of posting an untrue online review decreases. As the level of income increases, the log odds of posting an untrue online decrease.

**Table 9**

*Logistic regression model of respondents posting untrue online reviews (dependent variable: everlied)*

|  | Estimate | Std. Error | z value | Pr(|z|) | Significance |
|---|---|---|---|---|---|
| Intercept | -7.327 | 1.140 | -6.426 | <0.001 | (***) |
| Gender | 0.753 | 0.346 | 2.179 | 0.029 | (*) |
| Income | -0.429 | 0.175 | -2.458 | 0.018 | (*) |
| Region | -0.059 | 0.087 | -0.681 | 0.459 | |
| Education | 0.837 | 0.187 | 4.489 | <0.001 | (***) |
| Age | -0.477 | 0.201 | -2.374 | 0.018 | (*) |
| Total People Trust Score | 0.131 | 0.022 | 5.904 | <0.001 | (***) |

[*]*p* < .05. [**]*p* < .01. [***]*p* < .001.

**Source:** Data analysis, Berry (2024)

## 4. Discussion

The findings suggest that the profile of the majority of online users that tend to post untrue reviews can be generally characterized as being male, earning less than $49,999 per year, holding a bachelor's or master's degree, residing in the states of New York, New Jersey and Pennsylvania, in the age range of 25 to 44 years, and largely appear to neither trust nor distrust religious leaders, business owners, new immigrants, and social media influencers, among others, implying ambivalent attitudes toward certain people. Furthermore, many of the respondents that admitted to posting untrue online reviews appear to harbor strong distrust of celebrities, government officials, new immigrants, and salespeople, perhaps suggesting a disdain for certain authority individuals or those that are in a position to influence the public in some way through their popularity.

As a global proxy for lying online, the findings suggest that almost 17% of people post untrue information, which falls into the range observed by Wu et al. (2020). Additionally, based on the observation of 20 of 44 respondents (or about 5% of the original n=398 panel size) that were excluded from the sample for trying to defeat location screening and gaming the system to be compensated more than once on Amazon Mechanical Turk, implying that some mTurk users were willing to be dishonest for financial gain, and at the expense of data quality. These findings are problematic for not just businesses and consumers but also academic researchers who must clean data, and estimate the extent to which a data set may be potentially compromised by fakery. However, the results point to the fact that, despite using adequate screening questions and being assured that the user population being sampled is from a given country and age group, global survey respondents have somehow found a way to exploit the weaknesses in mTurk's quality control system. Therefore, based on the findings from this study regarding excluded respondents, researchers using mTurk should expect to 11-12% of a desired sample size to be excluded for various reasons. Researchers should perhaps increase their sample size by this percentage to buffer the possible reduction in usable data.

The logistic regression model suggests that the level of education and trust of people score are highly statistically significant predictors of posting untrue online reviews, confirming the observation of Brühlmann et al. (2020). The region of residence in the United States does not appear to be a significant predictor. The finding that age is a statistically significant predictor, confirming the claim of Ariely et al. (2019), and also confirms the claim of Gerlach, Teodorescu and Hertwig (2019) that younger people tend to lie more than older people. While the level of income of respondents was a statistically significant predictor, the findings did not bear out the claim of Piff et al. (2012) that upper-class individuals were more unethical than lower-class individuals. Indeed, those respondents at lower income levels appeared to post untrue reviews more than those at higher income levels. Finally, gender, is a statistically significant predictor, and confirms the finding by Gerlach, Teodorescu and Hertwig (2019) that men generally lie more than women, given the log odds increase in likelihood from gender=0 for females to gender=1 for men.

### 5. Directions for future research

First, future research should attempt to leverage the information that can be gleaned from non-compliant survey respondents to quantify different kinds of deception by respondents as this is highly relevant to data quality, such as attention check failures and trying to subvert questions screener controls. Second, scholars should incorporate the trust of people instrument from Berry (2024) as a reliable means of evaluating the extent to which people are not just trusting but also as potential cues for general hostility or antisocial behaviors among people. Third, these findings should help those studying the propensity for people to deliberately spread misinformation in different online venues. Finally, this study can serve as a basis for wider application to the study of dishonesty in many different settings, including justice, business, and education.

## 6. Limitations

This study, as with any given study, is subject to limitations that must be discussed. First, the author acknowledges that Amazon Mechanical Turk, while convenient, has well-documented shortcomings that can present challenges for researchers. Although best practices are used to ensure that questions are completed (Iowa State University, n.d.), researchers must attempt to include a variety of attention check questions that will defeat the possibility of lazy respondents or even bots. Second, the sample population of Amazon Mechanical Turk is somewhat problematic in that there is a subset of users claiming to be from the United States but located elsewhere. Although there is a limit to the kinds of personal information that may help to categorize respondents, verifying information about the true location of a user without breaching privacy concerns is required. Finally, while the demographic composition of Amazon Mechanical Turk users is somewhat different from the general population, the results of this study should still be considered generalizable, nonetheless.

## 7. Patents

There are no patents resulting from the work reported in this manuscript.

## 8. Funding

This research received no external funding.

## 9. Conflicts of Interest

The authors declare no conflict of interest.

## 10. Declaration of generative AI in scientific writing

The author declares that generative AI tools were not used in the writing or research of this article.